# An Adaptive Conditional Zero-Forcing Decoder with Full-diversity, Least Complexity and Essentially-ML Performance for STBCs

Lakshmi Prasad Natarajan and B. Sundar Rajan
Dept. of ECE, IISc, Bangalore 560012, India
Email: {nlp,bsrajan}@ece.iisc.ernet.in

*Abstract*—A low complexity, essentially-ML decoding technique for the Golden code and the 3 antenna Perfect code was introduced by Sirianunpiboon, Howard & Calderbank. Though no theoretical analysis of the decoder was given, the simulations showed that this decoding technique has almost maximum-likelihood (ML) performance. Inspired by this technique, in this paper we introduce two new low complexity decoders for Space-Time Block Codes (STBCs) - the Adaptive Conditional Zero-Forcing (ACZF) decoder and the ACZF decoder with successive interference cancellation (ACZF-SIC), which include as a special case the decoding technique of Sirianunpiboon et al. We show that both ACZF and ACZF-SIC decoders are capable of achieving full-diversity, and we give sufficient conditions for an STBC to give full-diversity with these decoders. We then show that the Golden code, the 3 and 4 antenna Perfect codes, the 3 antenna Threaded Algebraic Space-Time code and the 4 antenna rate 2 code of Srinath & Rajan are all full- diversity ACZF/ACZF-SIC decodable with complexity strictly less than that of their ML decoders. Simulations show that the proposed decoding method performs identical to ML decoding for all these five codes. These STBCs along with the proposed decoding algorithm outperform all known codes in terms of decoding complexity and error performance for $N_t \leq 4$ transmit antennas. We further provide a lower bound on the complexity of full-diversity ACZF/ACZF-SIC decoding. All the five codes listed above achieve this lower bound and hence are optimal in terms of minimizing the ACZF/ACZF-SIC decoding complexity. Both ACZF and ACZF-SIC decoders are amenable to sphere decoding implementation.

## I. Introduction

The problem of constructing Space-Time Block Codes (STBCs) that provide good error performance with low complexity decoding has drawn much attention in the literature. Low complexity decoding techniques such as zero-forcing (ZF), Minimum Mean Squared Error (MMSE) [1] and Partial Interference Cancellation [2] are capable of achieving full-diversity, but their error performance is considerably inferior to that of maximum-likelihood (ML) decoding. Consequently much work has been directed towards designing full-diversity codes that admit low complexity decoding with ML performance [3]–[16].

We now briefly review the best known full-diversity, low ML decoding complexity, high-rate STBCs for $N_t \leq 4$ transmit antennas. The Silver code [10], [11] has the least known ML decoding complexity of $M^2$ (where $M$ is the size of the complex constellation used) among known full-rate full-diversity codes for 2 transmit antennas. This is followed by the Golden code [17], [18] that has a higher ML decoding complexity $M^{2.5}$ [12], [13], but has superior coding gain and error performance than the Silver code. For 3 antenna systems the full-rate, full-diversity code with least ML decoding complexity is the Threaded Algebraic Space-Time (TAST) code [19] with a complexity of $M^7$ [14], whereas the code with the best known coding gain is the $3 \times 3$ Perfect code [20] that has a complexity of $M^9$. For 4 antenna systems, the $4 \times 4$ Perfect code has the best coding gain and an ML decoding complexity of $M^{13.5}$ [21]. Among the rate 2 codes for asymmetric MIMO systems with 4 transmit and 2 receive antennas the Srinath-Rajan code [12] has the least ML decoding complexity $M^{4.5}$ and best error performance.

It is possible to reduce the decoding complexity further by using a non-ML decoder without trading off the error performance, unlike ZF, MMSE or Partial Interference Cancellation receivers where the decoding comfort is achieved at the cost of higher probability of error. Such a decoding technique was proposed by Sirianunpiboon, Howard & Calderbank in [15], [16] for the Golden code and the 3 antenna Perfect code. Though no theoretical analysis of the achievable diversity or coding gain was provided, the simulations showed that these decoders have essentially the same performance as an ML decoder for the Golden and the three antenna Perfect code, but with complexity less than that of ML decoding.

The contributions and organization of this paper are as follows.

- Inspired by [15], [16], we introduce two new low complexity decoding algorithms for STBCs - the Adaptive Conditional Zero-Forcing (ACZF) decoder and the ACZF decoder with successive interference cancellation (ACZF-SIC) (Sections II-B). We show that these decoders are capable of achieving full-diversity in wireless Rayleigh faded channels and give sufficient conditions for an STBC to give full-diversity with ACZF/ACZF-SIC decoding (Sections III). The proposed decoders include as special case the decoding technique of [15], [16] for the Golden and 3 antenna Perfect code.
- We show that the best known codes for 2, 3, 4 antennas: the Perfect codes for 2, 3, 4 antennas, the 3 antenna TAST code and the Srinath-Rajan code, are all full-diversity ACZF/ACZF-SIC decodable with complexity strictly less than their ML decoding complexity (Section IV). See

## TABLE I
### DELAY-OPTIMAL CODES WITH LOW COMPLEXITY, FULL-DIVERSITY DECODING ALGORITHMS

| Code | Transmit Antennas $N_t$ | Rate $R$ | ML Decoding Complexity | **ACZF/ACZF-SIC Decoding Complexity** | Full-diversity codes with least known ML decoding complexity ||
|---|---|---|---|---|---|---|
| | | | | | ML Decoding Complexity | Code |
| Golden | 2 | 2 | $M^{2.5}$ | $\mathbf{M^2}$ | $M^2$ | Silver [10], [11] |
| Perfect | 3 | 3 | $M^9$ | $\mathbf{M^7}$ | $M^7$ | TAST |
| TAST | 3 | 3 | $M^7$ | $\mathbf{M^6}$ | $M^7$ | TAST |
| Perfect | 4 | 4 | $M^{13.5}$ | $\mathbf{M^{12}}$ | $M^{13.5}$ | Perfect |
| Srinath-Rajan | 4 | 2 | $M^{4.5}$ | $\mathbf{M^4}$ | $M^{4.5}$ | Srinath-Rajan |

$M$ is the size of the underlying complex constellation.

- Table I for comparison of ACZF/ACZF-SIC and ML decoding complexities of these codes.
- Simulation results (Section VI) show that the proposed decoder performs identical to the ML decoder for all these five codes, i.e., reduction in decoding complexity is achieved without trading off error performance. Thus these STBCs along with the proposed decoding algorithm outperform all known codes in terms of decoding complexity and error performance for $N_t \leq 4$. In particular, the Golden code outperforms the Silver code in both decoding complexity and error performance. See Table I for a comparison of the complexities of known full-diversity low decoding complexity codes.
- We derive a lower bound on the complexity of full-diversity ACZF/ACZF-SIC decoding (Section III-B). All the five codes mentioned above achieve this lower bound and hence are optimal in terms of minimizing the ACZF/ACZF-SIC decoding complexity.
- Both ACZF and ACZF-SIC algorithms are amenable to sphere decoding [22] implementation. We show that the ACZF-SIC decoder can be implemented with only a few minor modifications to the original sphere decoding algorithm (Section V).

The channel model is discussed in Section II-A and the paper is concluded in Section VII.

**Notation:** Matrices (vectors) are denoted by bold, uppercase (lowercase) letters. The Hermitian, transpose and Frobenius norm of a matrix $\mathbf{X}$ are denoted by $\mathbf{X}^H$, $\mathbf{X}^T$ and $||\mathbf{X}||$ respectively. The determinant of a square matrix $\mathbf{X}$ is denoted by $\det(\mathbf{X})$. For any vector $\mathbf{u}$ the diagonal matrix with the elements of $\mathbf{u}$ on the main diagonal is denoted by $\text{diag}(\mathbf{u})$. Unless used as a subscript or to denote indices, $j$ represents $\sqrt{-1}$. For any set $\mathcal{I}$, its complement in the corresponding universal set is denoted by $\mathcal{I}^c$. The expectation operator is denoted by $\mathbb{E}(\cdot)$ and the probability of an event $\mathcal{E}$ is denoted by $\mathsf{P}(\mathcal{E})$. For any vector $\mathbf{u}$, its $\ell^{th}$ component is denoted by $\mathbf{u}(\ell)$. The nearest integer operator is denoted by $\text{rnd}(\cdot)$. The notation $\mathbf{0}$ represents the all zero matrix of the appropriate dimension. For any matrix $A$ let $\text{vec}(A)$ denote the vectorization of $A$, i.e., the vector obtained by stacking the columns of $A$ one below another.

## II. ADAPTIVE CONDITIONAL ZERO-FORCING DECODER

We first explain the MIMO channel model used in this paper and then introduce the ACZF/ACZF-SIC decoding of STBCs.

### A. Channel Model

We consider a quasi-static Rayleigh flat-fading channel

$$\mathbf{Y} = \sqrt{\mathsf{SNR}}\mathbf{XH} + \mathbf{N} \quad (1)$$

with $N_t$ transmit antennas, $N_r$ receive antennas and delay $T$. The transmit matrix $\mathbf{X}$ takes values from a Space-Time Block Code (STBC) $\mathcal{C}$ which is a finite subset of $\mathbb{C}^{T \times N}$. The $N_t \times N_r$ channel matrix $\mathbf{H}$ is known at the receiver but not at the transmitter. The entries of $\mathbf{H}$ and the noise matrix $\mathbf{N}$ are assumed to be independent and identically distributed, zero mean circularly symmetric complex Gaussian random variables with unit variance. The transmit matrix $\mathbf{X}$ satisfies $\mathbb{E}\left(||\mathbf{X}||^2\right) = T$, so that the average signal to noise ratio at each receive antenna is equal to SNR. It is assumed that the STBC $\mathcal{C}$ is obtained via a *design* [23] $\mathbf{S} = \sum_{i=1}^{K} s_i \mathbf{A}_i$, where $\mathbf{A}_i \in \mathbb{C}^{T \times N}$, $i = 1, \ldots, K$, are the *linear dispersion* or *weight matrices*, and $s_1, \ldots, s_K$ are complex symbols that assume values from a finite constellation $\mathcal{A} \subset \mathbb{C}$, i.e.,

$$\mathcal{C} = \left\{ \sum_{i=1}^{K} s_i \mathbf{A}_i \ \Big|\ s_i \in \mathcal{A},\ i = 1, \ldots, K \right\}.$$

The signal set $\mathcal{A}$ is usually (but not always) a QAM, HEX or PSK constellation. The rate of $\mathcal{C}$ is $R = \frac{K}{T}$ in terms of complex symbols per channel use, and $\frac{K \log_2 |\mathcal{A}|}{T}$ in terms of bits per channel use.

Vectorizing the receive matrix $\mathbf{Y}$ in (1) we obtain

$$\mathbf{y} = \sqrt{\mathsf{SNR}}\mathbf{Gs} + \mathbf{n},$$

where $\mathbf{y}$ and $\mathbf{n}$ are vectorizations of $\mathbf{Y}$ and $\mathbf{N}$ respectively, $\mathbf{s} = [s_1\ s_2 \cdots s_K]^T$, and the *equivalent channel matrix* $\mathbf{G} \in \mathbb{C}^{N_r T \times K}$ is a function of the channel $\mathbf{H}$ and the design $\mathbf{S}$ and is given by $[\text{vec}(\mathbf{A}_1\mathbf{H})\ \text{vec}(\mathbf{A}_2\mathbf{H}) \cdots \text{vec}(\mathbf{A}_K\mathbf{H})]$.

*Example G.1:* The Golden Code [20] is a rate $R = 2$ code for $N_t = 2$ antennas. It encodes $K = 4$ symbols from a QAM alphabet over $T = 2$ time slots. The design is

$$\begin{bmatrix} s_1 & js_2 \\ s_2 & s_1 \end{bmatrix} \begin{bmatrix} \alpha & 0 \\ 0 & \bar{\alpha} \end{bmatrix} + \begin{bmatrix} s_3 & js_4 \\ s_4 & s_3 \end{bmatrix} \begin{bmatrix} \alpha\tau & 0 \\ 0 & \bar{\alpha}\mu \end{bmatrix},$$

where $\tau = \frac{1+\sqrt{5}}{2}$, $\mu = -\frac{1}{\tau}$, $\alpha = 1 + j\mu$ and $\bar{\alpha} = 1 + j\tau$. The four weight matrices are

$$\mathbf{A}_1 = \begin{bmatrix} \alpha & 0 \\ 0 & \bar{\alpha} \end{bmatrix}, \quad \mathbf{A}_2 = \begin{bmatrix} 0 & j\bar{\alpha} \\ \alpha & 0 \end{bmatrix},$$
$$\mathbf{A}_3 = \begin{bmatrix} \alpha\tau & 0 \\ 0 & \bar{\alpha}\mu \end{bmatrix} \text{ and } \mathbf{A}_4 = \begin{bmatrix} 0 & j\bar{\alpha}\mu \\ \alpha\tau & 0 \end{bmatrix}.$$

Consider the case $N_r = 2$ and let the channel matrix $\mathbf{H} = \begin{bmatrix} h_1 & h_2 \\ h_3 & h_4 \end{bmatrix}$. Then the equivalent channel $\mathbf{G} = [\text{vec}(\mathbf{A}_1\mathbf{H}) \ \text{vec}(\mathbf{A}_2\mathbf{H}) \cdots \text{vec}(\mathbf{A}_4\mathbf{H})]$ is given by

$$\begin{bmatrix} \alpha h_1 & j\bar{\alpha}h_3 & \alpha\tau h_1 & j\bar{\alpha}\mu h_3 \\ \bar{\alpha}h_3 & \alpha h_1 & \bar{\alpha}\mu h_3 & \alpha\tau h_1 \\ \alpha h_2 & j\bar{\alpha}h_4 & \alpha\tau h_2 & j\bar{\alpha}\mu h_4 \\ \bar{\alpha}h_4 & \alpha h_2 & \bar{\alpha}\mu h_4 & \alpha\tau h_2 \end{bmatrix}. \quad (2)$$

■

### B. Adaptive Conditional Zero-Forcing Decoder

We will introduce some notations before explaining the ACZF decoder. Let $\mathcal{I}_1, \ldots, \mathcal{I}_L$ be any $L$ subsets of $\{1, \ldots, K\}$ each of cardinality $\lambda$. The subsets $\mathcal{I}_1, \ldots, \mathcal{I}_L$ need not be a partition of $\{1, \ldots, K\}$, and they may have non-trivial intersections also. For any $\mathcal{I} \subseteq \{1, \ldots, K\}$ denote by $\mathbf{s}_\mathcal{I}$ the vector comprising of those symbols $s_i$ whose indices belong to $\mathcal{I}$, i.e., if $\mathcal{I} = \{i_1, i_2, \ldots, i_{|\mathcal{I}|}\}$ with $i_1 < i_2 < \cdots < i_{|\mathcal{I}|}$ then $\mathbf{s}_\mathcal{I} = [s_{i_1} s_{i_2} \cdots s_{i_{|\mathcal{I}|}}]^T$. Similarly let $\mathbf{G}_\mathcal{I}$ be the submatrix of $\mathbf{G}$ comprising of those columns whose indices belong to $\mathcal{I}$, i.e., $[\text{vec}(\mathbf{A}_{i_1}\mathbf{H}) \ \text{vec}(\mathbf{A}_{i_2}\mathbf{H}) \cdots \text{vec}(\mathbf{A}_{i_{|\mathcal{I}|}}\mathbf{H})]$. Further, let the operator $(\cdot)^\dagger$ denote the left pseudo-inverse of a matrix, and $\mathbf{u}(j)$ denote the $j^{th}$ entry of a vector $\mathbf{u}$.

The ACZF decoder, given in Algorithm 1, functions as follows. Given $\mathbf{y}$ and $\mathbf{G}$, from among $\mathcal{I}_1, \ldots, \mathcal{I}_L$ the subset $\mathcal{I}_m$ with the maximum value of $\det(\mathbf{G}_{\mathcal{I}_m}^H \mathbf{G}_{\mathcal{I}_m})$ is chosen. For each of the possible $|\mathcal{A}|^{K-\lambda}$ values of the vector $\mathbf{s}_{\mathcal{I}_m^c}$, the interference in $\mathbf{y}$ from $\mathbf{s}_{\mathcal{I}_m^c}$ is first removed and the remaining symbols $\mathbf{s}_{\mathcal{I}_m}$ for this instantiation of $\mathbf{s}_{\mathcal{I}_m^c}$ are decoded by zero-forcing resulting in the decoded vector $\hat{\mathbf{s}}_{\mathcal{I}_m}(\mathbf{s}_{\mathcal{I}_m^c})$. At the end of this step we have a set of $|\mathcal{A}|^{K-\lambda}$ tuples

$$\left\{ \left(\mathbf{s}_{\mathcal{I}_m^c}, \hat{\mathbf{s}}_{\mathcal{I}_m}(\mathbf{s}_{\mathcal{I}_m^c})\right) \ \middle| \ \mathbf{s}_{\mathcal{I}_m^c} \in \mathcal{A}^{|\mathcal{I}_m^c|} \right\}. \quad (3)$$

From among these vectors the decoder chooses the tuple $\hat{\mathbf{s}}$ that minimizes $||\mathbf{y} - \mathbf{G}\mathbf{s}||^2$.

After the interference from $\mathbf{s}_{\mathcal{I}_m^c}$ is removed from $\mathbf{y}$ the symbols in $\mathbf{s}_{\mathcal{I}_m}$ are decoded by a ZF receiver. The post-processing signal to noise ratio for decoding $\mathbf{s}_{\mathcal{I}_m}$ is captured by $\det(\mathbf{G}_{\mathcal{I}_m}^H \mathbf{G}_{\mathcal{I}_m})$ [15], [16]. To optimize the error performance the decoder therefore chooses the subset whose equivalent channel matrix has the largest determinant.

*Example G.2:* Continuing Example G.1, consider the ACZF decoder for the Golden code with parameters $L = 2$, $\mathcal{I}_1 = \{1, 2\}$ and $\mathcal{I}_2 = \{3, 4\}$. In this case $\lambda = 2$, $\mathbf{s}_{\mathcal{I}_1} = [s_1 \ s_2]^T$, $\mathbf{s}_{\mathcal{I}_2} = [s_3 \ s_4]^T$, $\mathbf{G}_{\mathcal{I}_1}$ comprises of the first two columns of $\mathbf{G}$ in (2) and $\mathbf{G}_{\mathcal{I}_2}$ comprises of the last two columns. The decoder first determines $m = \arg\max_{\ell \in \{1,2\}} \det(\mathbf{G}_{\mathcal{I}_\ell}^H \mathbf{G}_{\mathcal{I}_\ell})$. Note that in this case

---

**Input**: Received signal $\mathbf{y}$, equivalent channel matrix $\mathbf{G}$, subsets $\mathcal{I}_1, \ldots, \mathcal{I}_L$
**Output**: Decoded symbol vector $\hat{\mathbf{s}}$
Set $m := 0$, maxdet $:= 0$;
**foreach** $\ell = 1, \ldots, L$ **do**
    **if** maxdet $< \det(\mathbf{G}_{\mathcal{I}_\ell}^H \mathbf{G}_{\mathcal{I}_\ell})$ **then**
        maxdet $:= \det(\mathbf{G}_{\mathcal{I}_\ell}^H \mathbf{G}_{\mathcal{I}_\ell})$;
        $m := \ell$;
    **end**
**end**
Set $\mathbf{u}$ as the all zero vector of length $\lambda$;
**foreach** $\mathbf{a}_{\mathcal{I}_m^c} \in \mathcal{A}^{K-\lambda}$ **do**
    Calculate $\mathbf{v} := \frac{1}{\sqrt{\text{SNR}}} \mathbf{G}_{\mathcal{I}_m}^\dagger (\mathbf{y} - \sqrt{\text{SNR}} \mathbf{G}_{\mathcal{I}_m^c} \mathbf{a}_{\mathcal{I}_m^c})$;
    **foreach** $j = 1, \ldots, \lambda$ **do**
        Find $\mathbf{u}(j) := \arg\min_{a \in \mathcal{A}} ||\mathbf{v}(j) - a||$;
    **end**
    Set $\hat{\mathbf{s}}_{\mathcal{I}_m}(\mathbf{a}_{\mathcal{I}_m^c}) := \mathbf{u}$;
**end**
Set mindis $:= \infty$ and $\hat{\mathbf{s}}_{\mathcal{I}_m^c} := \mathbf{0}$;
**foreach** $\mathbf{a}_{\mathcal{I}_m^c} \in \mathcal{A}^{K-\lambda}$ **do**
    **if**
    mindis $> ||\mathbf{y} - \sqrt{\text{SNR}}(\mathbf{G}_{\mathcal{I}_m^c} \mathbf{a}_{\mathcal{I}_m^c} - \mathbf{G}_{\mathcal{I}_m} \hat{\mathbf{s}}_{\mathcal{I}_m}(\mathbf{a}_{\mathcal{I}_m^c}))||^2$
    **then**
        mindis $:=$
        $||\mathbf{y} - \sqrt{\text{SNR}}(\mathbf{G}_{\mathcal{I}_m^c} \mathbf{a}_{\mathcal{I}_m^c} - \mathbf{G}_{\mathcal{I}_m} \hat{\mathbf{s}}_{\mathcal{I}_m}(\mathbf{a}_{\mathcal{I}_m^c}))||^2$;
        $\hat{\mathbf{s}}_{\mathcal{I}_m^c} := \mathbf{a}_{\mathcal{I}_m^c}$;
    **end**
**end**
Set $\hat{\mathbf{s}}$ as the concatenation of $\hat{\mathbf{s}}_{\mathcal{I}_m^c}$ and $\hat{\mathbf{s}}_{\mathcal{I}_m}(\hat{\mathbf{s}}_{\mathcal{I}_m^c})$;

**Algorithm 1:** The ACZF Decoder.

---

$\mathcal{I}_m^c = \mathcal{I}_{2-m}$. Conditioned on $\mathbf{s}_{\mathcal{I}_{2-m}}$ the receiver decodes $\hat{\mathbf{s}}_{\mathcal{I}_m}(\mathbf{s}_{\mathcal{I}_{2-m}})$ by zero-forcing. Then from among the $M^2$ tuples $\left\{ \left(\mathbf{s}_{\mathcal{I}_{2-m}}, \hat{\mathbf{s}}_{\mathcal{I}_m}(\mathbf{s}_{\mathcal{I}_{2-m}})\right) \right\}$ the one that minimizes $||\mathbf{y} - \mathbf{G}\mathbf{s}||^2$ is chosen to be the output. This decoder for the Golden Code was first proposed in [16], but no theoretical analysis regarding diversity or minimum achievable complexity was provided. ■

It is well known [26] that at high values of SNR the ZF receiver aided with successive interference cancellation (ZF-SIC) performs better than mere ZF decoding. The ACZF decoder can be integrated with SIC (ACZF-SIC) as well. The ACZF-SIC decoder detects the symbols in $\mathbf{s}_{\mathcal{I}_m}$ one by one, and removes the effect of the already detected symbols in the received vector by interference cancellation.

*Complexity Analysis:* Consider the step $\mathbf{u}(j) = \arg\min_{a \in \mathcal{A}} ||\mathbf{v}(j) - a||$ in Algorithm 1. If the size of the complex constellation $\mathcal{A}$ is $M$ this step requires $M$ computations of $||\mathbf{v}(j) - a||$. However if $\mathcal{A}$ is a regular $M$-ary QAM constellation then this step can be implemented with constant complexity independent of $M$ by hard limiting [8], [12], [13] as shown in (4) and (5) at the top of the next page, where $(\cdot)_{Re}$ and $(\cdot)_{Im}$ denote the real and imaginary

$$(\mathbf{u}(j))_{Re} = \min\left\{\max\left\{\mathsf{rnd}\left(\frac{\sqrt{M}-1}{2} - (\mathbf{v}(j))_{Re}\right), 0\right\}, \sqrt{M}-1\right\} - \frac{\sqrt{M}-1}{2}. \quad (4)$$

$$(\mathbf{u}(j))_{Im} = \min\left\{\max\left\{\mathsf{rnd}\left(\frac{\sqrt{M}-1}{2} - (\mathbf{v}(j))_{Im}\right), 0\right\}, \sqrt{M}-1\right\} - \frac{\sqrt{M}-1}{2}. \quad (5)$$

parts respectively. This step is performed $|\lambda|M^{K-\lambda}$ times throughout the decoding process. Thus the complexity order of the ACZF decoder is $M^{K-\lambda}$ and $M^{K-\lambda+1}$ for QAM and general constellations respectively. Identical results hold true for the complexity of ACZF-SIC decoding also. For example, for the ACZF decoder for the Golden Code in Example G.2 $K = 4$, $\lambda = 2$ and $\mathcal{A}$ is a regular QAM constellation. Hence this decoder has a complexity of $M^{K-\lambda} = M^2$. On the other hand the least known ML decoding complexity of the Golden Code is $M^{2.5}$ [12], [13].

## III. FULL-DIVERSITY CRITERION

In this section we give two equivalent full-diversity criteria for ACZF/ACZF-SIC decoders. We then give a lower bound on the complexity of full-diversity ACZF/ACZF-SIC decoding.

### A. Full-diversity Criterion

The vector $\mathbf{v}$ in Algorithm 1 is obtained by multiplying another vector by the left pseudo-inverse of $\mathbf{G}_{\mathcal{I}_m}$. Thus it is implicitly assumed that $\mathbf{G}_{\mathcal{I}_m}$ has full column rank. Note that this matrix is the choice from among $\mathbf{G}_{\mathcal{I}_1}, \cdots, \mathbf{G}_{\mathcal{I}_L}$ with the largest determinant. The ACZF decoding algorithm thus assumes that for every channel realization $\mathbf{H} \neq \mathbf{0}$ at least one of the $L$ matrices $\mathbf{G}_{\mathcal{I}_1}, \cdots, \mathbf{G}_{\mathcal{I}_L}$ has full column rank. The same is true in the case of ACZF-SIC decoding also. In the following theorem we show that this condition, which is necessary for the implementation of ACZF/ACZF-SIC decoder, is also a sufficient condition for ACZF to achieve the same diversity as the ML decoder.

*Theorem 1:* The ACZF decoder achieves the same diversity order as the ML decoder if for every channel realization $\mathbf{H} \neq \mathbf{0}$ at least one of the $L$ matrices $\mathbf{G}_{\mathcal{I}_1}, \cdots, \mathbf{G}_{\mathcal{I}_L}$ has full column rank.

*Proof:* Proof is given in Appendix A ∎

Both zero-forcing and conditional zero-forcing decoders [24] are special cases of ACZF decoding. In both these cases $L = 1$ and hence the decoder is not 'adaptive' any more, i.e., for every channel $\mathbf{H}$ we have $m = 1$. When the number of subsets $L = 1$ and $\mathcal{I}_1 = \{1, \ldots, K\}$ the ACZF decoder reduces to the zero-forcing receiver. The criterion of Theorem 1 in this case reduces to the full-diversity criterion for ZF decoding given in [1], [2]. If $L = 1$ and $\mathcal{I}_1 \subsetneq \{1, \ldots, K\}$ then the ACZF decoder reduces to conditional ZF decoder [24]. In this case Theorem 1 implies that $\mathbf{G}_{\mathcal{I}_1}$ be of full column rank for every $\mathbf{H} \neq \mathbf{0}$ and the difference of any two codewords in $\mathcal{C}$ be of rank $N_t$ for the conditional ZF decoder to achieve a diversity order of $N_t N_r$.

This coincides with the full-diversity criterion for conditional ZF decoders given in [24].

Theorem 1 imposes a criterion on the equivalent channel matrix $\mathbf{G}$ which is a function of the linear dispersion matrices and the number of receive antennas. In the following theorem we give an equivalent full-diversity criterion in terms of the linear dispersion matrices alone. This criterion shows that the sufficient condition in Theorem 1 is independent of the number of receive antennas, i.e., either the STBC $\mathcal{C}$ satisfies the criterion for all $N_r \geq 1$ or it does not satisfy the criterion for any $N_r$. We now introduce some notations towards stating the equivalent criterion. For any $\mathcal{I} = \{i_1, \ldots, i_{|\mathcal{I}|}\}$ and any vector $\mathbf{u} = [u_1 \, u_2 \, \cdots \, u_{|\mathcal{I}|}] \in \mathbb{C}^{|\mathcal{I}|}$ let $\mathbf{X}_{\mathcal{I}}(\mathbf{u}) = \sum_{j=1}^{|\mathcal{I}|} u_j \mathbf{A}_{i_j}$, i.e., $\mathbf{X}_{\mathcal{I}}(\mathbf{u})$ is the complex linear combination of the weight matrices with indices in $\mathcal{I}$ with the elements of vector $\mathbf{u}$ defining the corresponding complex coefficients.

*Theorem 2:* The ACZF decoder achieves the same diversity order as the ML decoder if for every choice of $\mathbf{u}_1 \in \mathbb{C}^\lambda \setminus \{\mathbf{0}\}$, $\mathbf{u}_2 \in \mathbb{C}^\lambda \setminus \{\mathbf{0}\}, \ldots, \mathbf{u}_L \in \mathbb{C}^\lambda \setminus \{\mathbf{0}\}$, the $LT \times N_t$ matrix

$$\widetilde{\mathbf{X}}(\mathbf{u}_1, \ldots, \mathbf{u}_L) = \begin{bmatrix} \mathbf{X}_{\mathcal{I}_1}(\mathbf{u}_1) \\ \mathbf{X}_{\mathcal{I}_2}(\mathbf{u}_2) \\ \vdots \\ \mathbf{X}_{\mathcal{I}_L}(\mathbf{u}_L) \end{bmatrix} \quad (6)$$

has full column rank.

*Proof:* Proof is given in Appendix B. ∎

The criteria of Theorems 1 and 2 ensure full-diversity with ACZF-SIC decoders also.

*Lemma 1:* The ACZF-SIC decoder achieves the same diversity order as the ML decoder if the STBC satisfies the criteria of Theorems 1 or 2.

*Proof:* Proof is given in Appendix C. ∎

The sufficient condition of Theorem 2 is independent of the number of receive antennas $N_r$ and the choice of the constellation $\mathcal{A}$, and depends only on the weight matrices $\mathbf{A}_1, \ldots, \mathbf{A}_K$. The signal set $\mathcal{A}$ can be chosen such that the STBC gives full diversity with ML decoding.

*Corollary 1:* If the difference of any two codewords of $\mathcal{C}$ is of rank $N_t$, and if $\mathcal{C}$ satisfies the criterion of Theorem 1 or 2, then $\mathcal{C}$ achieves a diversity of $N_t N_r$ with ACZF and ACZF-SIC decoding.

### B. Lower Bound on Decoding Complexity

Consider any code for $N_t$ antennas with delay $T$ that satisfies the criteria of Theorems 1 and 2 for certain number of receive antennas $N_r$. Then the code satisfies the criterion of Theorem 1 for $N_r = 1$ as well. For such a code with $N_r = 1$, for every channel $\mathbf{H} \neq \mathbf{0}$ the $T \times \lambda$ matrix $\mathbf{G}_{\mathcal{I}_m}$ has full column rank. This implies that $\lambda \leq T$ and that the order of decoding complexity is at least $M^{K-T}$ and $M^{K-T+1}$ for QAM and arbitrary signal sets respectively. If we restrict our attention to minimum-delay codes that provide full diversity we have $T = N_t$, and the lower bound on decoding complexity is $M^{K-N_t}$ and $M^{K-N_t+1}$ for QAM and arbitrary constellations respectively.

## IV. FULL-DIVERSITY ACZF/ACZF-SIC DECODABLE CODES WITH OPTIMAL DECODING COMPLEXITY

In this section we show that some of the best codes known for $N_t = 2, 3$ and $4$ antennas are full-diversity ACZF/ACZF-SIC decodable. These include the Perfect Codes [20] for $2, 3$ and $4$ antennas (the two antenna Perfect Code being the Golden Code [17], [18]), the $3 \times 3$ Threaded Algebraic Space-Time Code (TAST) [19] and the Srinath-Rajan Code [12] which is the best known rate $2$ code for $4$ transmit and $2$ receive antenna MIMO systems. All these codes achieve the lower bound on the decoding complexity given in Section III-B. The ACZF decoding algorithm for the 3 antenna Perfect Code given in this section was first proposed in [15], but without the proof for full-diversity. The results of this section are summarized in Table I. The table includes the ML and ACZF/ACZF-SIC decoding complexity of these codes, and the details of the full-diversity codes for same values of $(N_t, R)$ that have the previously least known decoding complexity. In all the cases, the proposed decoders have the least complexity among all known full-diversity decoding methods.

### A. The Golden Code

In this subsection we show that the ACZF/ACZF-SIC decoder for the Golden Code given in Example G.2 achieves full diversity. From Theorem 2 we need to show that the matrix

$$\widetilde{\mathbf{X}} = \begin{bmatrix} \alpha s_1 & j\bar{\alpha}s_2 \\ \alpha s_2 & \bar{\alpha}s_1 \\ \alpha\tau s_3 & j\bar{\alpha}\mu s_4 \\ \alpha\tau s_4 & \bar{\alpha}\mu s_3 \end{bmatrix}$$

has linearly independent columns whenever $[s_1 \ s_2]^T, [s_3 \ s_4]^T \in \mathbb{C}^2 \setminus \{\mathbf{0}\}$.

*Proof of full-diversity:* Proof is by contradiction. Suppose $[s_1 \ s_2]^T$ and $[s_3 \ s_4]^T$ are non-zero and the columns of $\widetilde{\mathbf{X}}$ are linearly dependent. We will first argue that none of $s_1, \ldots, s_4$ is equal to zero. If $s_2 = 0$ then $s_1$ can not be zero, as this would make the vector $[s_1 \ s_2]^T = \mathbf{0}$. Thus, if $s_2 = 0$ the upper $2$ submatrix of $\widetilde{\mathbf{X}}$ is a diagonal full-ranked matrix, and thus $\widetilde{\mathbf{X}}$ is of rank 2. This negates our initial assumption, and hence $s_2 \neq 0$. Using similar argument we have $s_1, s_3, s_4 \neq 0$ as well.

Let $[a \ b]^T$ be a non-zero vector in the nullspace of $\widetilde{\mathbf{X}}$. Since neither columns of $\widetilde{\mathbf{X}}$ are zero we have $a, b \neq 0$. Now consider the first row of $\widetilde{\mathbf{X}}$. We have $\alpha s_1 a + j\bar{\alpha}s_2 b = 0$. This implies that $|\alpha s_1 a| = |\bar{\alpha}s_2 b|$, i.e., $\frac{|s_1|}{|s_2|} = \frac{|\alpha a|}{|\bar{\alpha}b|}$. Similarly from the second row we obtain $\frac{|s_1|}{|s_2|} = \frac{|\bar{\alpha}b|}{|\alpha a|}$. These two relations imply that

$$\frac{|\alpha|}{|\bar{\alpha}|} = \frac{|b|}{|a|}. \qquad (7)$$

Using similar argument with the last two rows of $\widetilde{\mathbf{X}}$ we obtain

$$\frac{|\alpha\tau|}{|\bar{\alpha}\mu|} = \frac{|b|}{|a|}. \qquad (8)$$

However (7) and (8) together imply that $|\tau| = |\mu|$ which is not true. Thus, by contradiction, we have shown that $\widetilde{\mathbf{X}}$ has linearly independent columns whenever $[s_1 \ s_2]^T$ and $[s_3 \ s_4]^T$ are non-zero. Since the Golden Code achieves full-diversity with the ML decoder, it achieves full-diversity with the ACZF/ACZF-SIC decoder as well. ∎

Thus the Golden code is full-diversity decodable with complexity $M^2$. Note that this meets the lower bound on the decoding complexity given in Section III-B. Further, this is a reduction by $M^{0.5}$ from the ML decoding complexity.

### B. The $3 \times 3$ Perfect Code

This full-diversity ML decodable STBC encodes $K = 9$ information symbols that assume values from a HEX constellation. The delay $T = 3$, and the ACZF/ACZF-SIC decoder employs $L = 3$ subsets $\mathcal{I}_1 = \{1, 2, 3\}$, $\mathcal{I}_2 = \{4, 5, 6\}$ and $\mathcal{I}_3 = \{7, 8, 9\}$. Let $\gamma = e^{j\frac{2\pi}{3}}$ and

$$\mathbf{U} = \begin{bmatrix} 0 & 0 & \gamma \\ 1 & 0 & 0 \\ 0 & 1 & 0 \end{bmatrix}.$$

The weight matrix of the $k^{th}$ symbol in $\ell^{th}$ group i.e. the weight matrix of the symbol $s_{3(\ell-1)+k}$ is $\mathbf{A}_{3(\ell-1)+k} = \mathbf{U}^{k-1}\mathbf{D}_\ell$, where $\mathbf{D}_1, \mathbf{D}_2, \mathbf{D}_3$ are diagonal matrices that are specified in [20].

*Proof of full-diversity:* For any choice of complex vectors $\mathbf{s}_{\mathcal{I}_1}, \mathbf{s}_{\mathcal{I}_2}, \mathbf{s}_{\mathcal{I}_3} \in \mathbb{C}^3 \setminus \{\mathbf{0}\}$ we need to show that the matrix

$$\widetilde{\mathbf{X}} = \begin{bmatrix} \sum_{k=1}^{3} s_k \mathbf{U}^{k-1}\mathbf{D}_1 \\ \sum_{k=1}^{3} s_{3+k} \mathbf{U}^{k-1}\mathbf{D}_2 \\ \sum_{k=1}^{3} s_{6+k} \mathbf{U}^{k-1}\mathbf{D}_3 \end{bmatrix}$$

is of rank 3. Since $\mathbf{U}$ is unitary it has an orthonormal set of eigenvectors and can be decomposed as $\mathbf{V}\mathbf{\Lambda}\mathbf{V}^H$, where the columns $\mathbf{v}_1, \mathbf{v}_2, \mathbf{v}_3$ of $\mathbf{V}$ are the eigenvectors and $\mathbf{\Lambda} = \mathrm{diag}(\sigma_1, \sigma_2, \sigma_3)$ is the diagonal matrix comprising of the eigenvalues. Therefore, $\mathbf{U}^{k-1} = \mathbf{V}\mathbf{\Lambda}^{k-1}\mathbf{V}^H$, $k = 1, 2, 3$, and the rank of $\widetilde{\mathbf{X}}$ and

$$\begin{bmatrix} \sum_{k=1}^{3} s_k \mathbf{\Lambda}^{k-1}\mathbf{V}^H\mathbf{D}_1 \\ \sum_{k=1}^{3} s_{3+k} \mathbf{\Lambda}^{k-1}\mathbf{V}^H\mathbf{D}_2 \\ \sum_{k=1}^{3} s_{6+k} \mathbf{\Lambda}^{k-1}\mathbf{V}^H\mathbf{D}_3 \end{bmatrix} \qquad (9)$$

are same. For $\ell = 1, 2, 3$ the matrix $\sum_{k=1}^{3} s_{3(\ell-1)+k}\mathbf{\Lambda}^{k-1} = \mathrm{diag}(z_{3(\ell-1)+1}, z_{3(\ell-1)+2}, z_{3(\ell-1)+3})$ is a diagonal matrix, where the vectors $[s_{3(\ell-1)+1} \ s_{3(\ell-1)+2} \ s_{3(\ell-1)+3}]^T$ and $[z_{3(\ell-1)+1} \ z_{3(\ell-1)+2} \ z_{3(\ell-1)+3}]^T$ are related as

$$\begin{bmatrix} z_{3(\ell-1)+1} \\ z_{3(\ell-1)+2} \\ z_{3(\ell-1)+3} \end{bmatrix} = \begin{bmatrix} 1 & \sigma_1 & \sigma_1^2 \\ 1 & \sigma_2 & \sigma_2^2 \\ 1 & \sigma_3 & \sigma_3^2 \end{bmatrix} \begin{bmatrix} s_{3(\ell-1)+1} \\ s_{3(\ell-1)+2} \\ s_{3(\ell-1)+3} \end{bmatrix}.$$

The three eigenvalues of $\mathbf{U}$ are all distinct and hence the above transformation matrix is Vandermonde and thus is invertible. Since $\mathbf{s}_{\mathcal{I}_\ell}$ is non-zero, the vector $[z_{3(\ell-1)+1}\ z_{3(\ell-1)+2}\ z_{3(\ell-1)+3}]^T$ is also non-zero, and hence at least one of its components is of non-zero value. For each $\ell = 1, 2, 3$ let $i_\ell \in \mathcal{I}_\ell$ be such that $z_{i_\ell} \neq 0$. The $3 \times 3$ submatrix of (9) comprising of the three rows corresponding to $z_{i_1}, z_{i_2}, z_{i_3}$ is

$$\begin{bmatrix} z_{i_1} & 0 & 0 \\ 0 & z_{i_2} & 0 \\ 0 & 0 & z_{i_3} \end{bmatrix} \begin{bmatrix} \mathbf{v}_{i_1}^H \mathbf{D}_1 \\ \mathbf{v}_{i_2-3}^H \mathbf{D}_2 \\ \mathbf{v}_{i_3-6}^H \mathbf{D}_3 \end{bmatrix},$$

and this matrix is full ranked whenever the matrix on the right hand side of the factorization above is full-ranked. By direct computation we have verified that for every choice of $i_1 \in \mathcal{I}_1$, $i_2 \in \mathcal{I}_2$ and $i_3 \in \mathcal{I}_3$ this matrix is indeed full-ranked. Thus, for every choice of $\mathbf{s}_{\mathcal{I}_1}, \mathbf{s}_{\mathcal{I}_2}, \mathbf{s}_{\mathcal{I}_3} \in \mathbb{C}^3 \setminus \{\mathbf{0}\}$ there exists a $3 \times 3$ submatrix of $\widetilde{\mathbf{X}}$ that is full-ranked and hence the columns of $\widetilde{\mathbf{X}}$ are linearly independent. This completes the proof. ∎

For the proposed ACZF/ACZF-SIC decoder for the $3 \times 3$ Perfect Code $\lambda = 3$. Since the constellation $\mathcal{A}$ is a HEX signal set the decoder complexity is $M^{K-\lambda+1} = M^7$.

### C. The $4 \times 4$ Perfect Code

This is a full-diversity ML decodable STBC with parameters $N_t = 4$, $T = 4$ and $K = 16$. The complex symbols are encoded using a regular QAM constellation. The proposed ACZF/ACZF-SIC decoder for this code employs $L = 4$ subsets $\mathcal{I}_1 = \{1, 2, 3, 4\}$, $\mathcal{I}_2 = \{5, 6, 7, 8\}$, ..., $\mathcal{I}_4 = \{13, 14, 15, 16\}$. Let

$$\mathbf{U} = \begin{bmatrix} 0 & 0 & 0 & j \\ 1 & 0 & 0 & 0 \\ 0 & 1 & 0 & 0 \\ 0 & 0 & 1 & 0 \end{bmatrix}.$$

For $1 \leq k, \ell \leq 4$, the weight matrix of the $k^{th}$ symbol in $\ell^{th}$ group i.e. the weight matrix of the symbol $s_{4(\ell-1)+k}$ is $\mathbf{A}_{4(\ell-1)+k} = \mathbf{U}^{k-1} \mathbf{D}_\ell$, where $\mathbf{D}_1, \mathbf{D}_2, \mathbf{D}_3, \mathbf{D}_4$, are diagonal matrices that are specified in [20].

The proof of full-diversity for ACZF/ACZF-SIC decoding for this code is similar to that of the $3 \times 3$ Perfect Code given in the previous subsection, and hence we avoid producing it here. The complexity of this decoder is $M^{12}$ and this achieves the lower bound of Section III-B. The complexity of ML decoding is however $M^{13.5}$ [21].

### D. The $3 \times 3$ TAST Code

The 3 antenna TAST code gives full-diversity with ML decoding and has parameters $T = 3$ and $K = 9$. The information symbols are encoded using QAM constellation. The design is given in (10) at the top of the next page, where $\gamma = e^{\frac{j\pi}{15}}$ and the $3 \times 3$ matrix $\mathbf{M} = [m_{i,j}]$ is the real rotation matrix from [25].

This code can be full-diversity ACZF/ACZF-SIC decoded using $L = 3$ subsets with $\mathcal{I}_1 = \{1, 2, 3\}$, $\mathcal{I}_2 = \{4, 5, 6\}$, $\mathcal{I}_3 = \{7, 8, 9\}$. The weight matrix of the $k^{th}$ symbol in the $\ell^{th}$ subset i.e., that of the symbol $s_{3(\ell-1)+k}$ is $\mathbf{U}^{k-1}\mathbf{D}_\ell$, where

$$\mathbf{U} = \begin{bmatrix} 0 & 0 & \gamma \\ \gamma & 0 & 0 \\ 0 & \gamma & 0 \end{bmatrix},$$

and $\mathbf{D}_\ell$ is the matrix obtained by diagonalizing the $\ell^{th}$ column of $\mathbf{M}$. The proof of full-diversity is similar to that of the $3 \times 3$ Perfect code and hence is omitted. This code achieves the minimum ACZF/ACZF-SIC decoding complexity of $M^6$. On the other hand, the ML decoding complexity of this code is $M^7$ [14].

### E. The Srinath-Rajan Code

Among all known rate 2 codes that enable reduced complexity ML decoding (complexity less than $M^K$) for 4 transmit antenna, 2 receive antenna asymmetric MIMO systems, the Srinath-Rajan code has the best error performance and least ML decoding complexity $M^{4.5}$. For 4 and 16 QAM constellations this code provides full-diversity with ML decoding. In this subsection we will show that this code is full-diversity ACZF/ACZF-SIC decodable with complexity of $M^4$. This is a reduction by a factor of $M^{0.5}$ from its ML decoding complexity.

We now introduce some notations. For any two complex numbers $a, b$ let

$$\mathbb{A}(a, b) = \begin{bmatrix} a & b \\ -b^* & a^* \end{bmatrix}$$

be the Alamouti matrix embedding the complex numbers $a$ and $b$. Note that $\mathbb{A}(a,b)$ is a scaled unitary matrix for every non-zero vector $[a\ b]^T \in \mathbb{C}^2$, and $\mathbb{A}(a,b)$ equals the all-zero matrix if and only if both $a, b = 0$. For any vector $\mathbf{u} \in \mathbb{C}^2$ we have $||\mathbb{A}(a,b)\mathbf{u}||^2 = (|a|^2 + |b|^2)||\mathbf{u}||^2$. Let $\gamma = e^{\frac{j\pi}{4}}$, $\theta = \frac{1}{2}\tan^{-1}(2)$, $c = \cos\theta$ and $s = \sin\theta$.

The parameters of the Srinath-Rajan code are $K = 8$ and $T = 4$ and its design is

$$\begin{bmatrix} c\mathbb{A}(s_1, s_2) & \gamma s\mathbb{A}(js_3, js_4) \\ \gamma c\mathbb{A}(s_3, s_4) & s\mathbb{A}(js_1, js_2) \end{bmatrix} + \begin{bmatrix} s\mathbb{A}(js_5, js_6) & \gamma c\mathbb{A}(s_7, s_8) \\ \gamma s\mathbb{A}(js_7, js_8) & c\mathbb{A}(s_5, s_6) \end{bmatrix}.$$

The proposed ACZF/ACZF-SIC decoder has $L = 2$ subsets with $\mathcal{I}_1 = \{1, 2, 3, 4\}$ and $\mathcal{I}_2 = \{5, 6, 7, 8\}$. Since $\lambda = N_t$ this decoder has the minimum achievable complexity $M^4$.

*Proof of full-diversity:* In order to use Theorem 2 we need to show that for any choice of $\mathbf{s}_{\mathcal{I}_1}, \mathbf{s}_{\mathcal{I}_2} \in \mathbb{C}^4 \setminus \{\mathbf{0}\}$, the matrix

$$\widetilde{\mathbf{X}} = \begin{bmatrix} c\mathbb{A}(s_1, s_2) & \gamma s\mathbb{A}(js_3, js_4) \\ \gamma c\mathbb{A}(s_3, s_4) & s\mathbb{A}(js_1, js_2) \\ s\mathbb{A}(js_5, js_6) & \gamma c\mathbb{A}(s_7, s_8) \\ \gamma s\mathbb{A}(js_7, js_8) & c\mathbb{A}(s_5, s_6) \end{bmatrix}$$

has full column rank. We will prove this by contradiction.

Suppose $\widetilde{\mathbf{X}}$ does not have full-column rank. We will first show that none of the component Alamouti blocks is identically zero. Say $s_3, s_4 = 0$, since $\mathbf{s}_{\mathcal{I}_1} \neq \mathbf{0}$, at least one of $s_1$ or $s_2$ is non-zero. Thus $\mathbb{A}(s_1, s_2)$ and $\mathbb{A}(js_1, js_2)$ are scaled unitary matrices, while $\mathbb{A}(js_3, js_4)$ and $\mathbb{A}(s_3, s_4)$ are zero.

$$\begin{bmatrix} s_1 & \gamma^2 s_3 & \gamma s_2 \\ \gamma s_2 & s_1 & \gamma^2 s_3 \\ \gamma^2 s_3 & \gamma s_2 & s_1 \end{bmatrix} \begin{bmatrix} m_{1,1} & 0 & 0 \\ 0 & m_{2,1} & 0 \\ 0 & 0 & m_{3,1} \end{bmatrix} + \begin{bmatrix} s_4 & \gamma^2 s_6 & \gamma s_5 \\ \gamma s_5 & s_4 & \gamma^2 s_6 \\ \gamma^2 s_6 & \gamma s_5 & s_4 \end{bmatrix} \begin{bmatrix} m_{1,2} & 0 & 0 \\ 0 & m_{2,2} & 0 \\ 0 & 0 & m_{3,2} \end{bmatrix} + \begin{bmatrix} s_7 & \gamma^2 s_9 & \gamma s_8 \\ \gamma s_8 & s_7 & \gamma^2 s_9 \\ \gamma^2 s_9 & \gamma s_8 & s_7 \end{bmatrix} \begin{bmatrix} m_{1,3} & 0 & 0 \\ 0 & m_{2,3} & 0 \\ 0 & 0 & m_{3,3} \end{bmatrix}.$$
(10)

Thus the upper $4 \times 4$ submatrix of $\widetilde{\mathbf{X}}$ is of full-rank, and hence $\widetilde{\mathbf{X}}$ has full column rank. Since this is a contradiction, at least one of $s_3, s_4$ is non-zero and hence $\mathbb{A}(js_3, js_4)$ and $\mathbb{A}(s_3, s_4)$ are non-zero matrices. Using similar arguments we can prove that all the Alamouti blocks in $\widetilde{\mathbf{X}}$ are non-zero.

Let $\mathbf{h} = [\mathbf{h}_1^T \ \mathbf{h}_2^T]^T$, with $\mathbf{h}_1, \mathbf{h}_2 \in \mathbb{C}^2$, be a non-zero vector in the null-space of $\widetilde{\mathbf{X}}$. Since the blocks have Alamouti structure, the first two columns and the last two columns of $\widetilde{\mathbf{X}}$ are orthogonal pairs. Thus, the submatrix of $\widetilde{\mathbf{X}}$ comprising the first two columns and the submatrix comprising the last two columns are both of rank 2. This implies that neither $\mathbf{h}_1$ nor $\mathbf{h}_2$ is a zero vector. Now consider the first two rows of $\widetilde{\mathbf{X}}$. Since $\mathbf{h}$ is in the null-space we have

$$c\mathbb{A}(s_1, s_2)\mathbf{h}_1 + \gamma s \mathbb{A}(js_3, js_4)\mathbf{h}_2 = \mathbf{0}.$$

In particular, $||c\mathbb{A}(s_1, s_2)\mathbf{h}_1||^2 = ||\gamma s \mathbb{A}(js_3, js_4)\mathbf{h}_2||^2$, i.e.,

$$c^2(|s_1|^2 + |s_2|^2)||\mathbf{h}_1||^2 = s^2(|s_3|^2 + |s_4|^2)||\mathbf{h}_2||^2. \quad (11)$$

Using the same technique on the second two rows of $\widetilde{\mathbf{X}}$

$$c^2(|s_3|^2 + |s_4|^2)||\mathbf{h}_1||^2 = s^2(|s_1|^2 + |s_2|^2)||\mathbf{h}_2||^2. \quad (12)$$

From (11) and (12) we have

$$\frac{||\mathbf{h}_1||}{||\mathbf{h}_2||} = \frac{s}{c}. \quad (13)$$

Repeating this method on the last four rows of $\widetilde{\mathbf{X}}$ we get

$$\frac{||\mathbf{h}_1||}{||\mathbf{h}_2||} = \frac{c}{s}. \quad (14)$$

However, (13) and (14) imply that $c = s$ which is not true. Thus $\widetilde{\mathbf{X}}$ has full column rank for any choice of $\mathbf{s}_{\mathcal{I}_1}, \mathbf{s}_{\mathcal{I}_2} \in \mathbb{C}^4 \setminus \{\mathbf{0}\}$. This completes the proof. ∎

## V. SPHERE DECODING IMPLEMENTATION

The sphere decoding algorithm [22] can be used to ML decode STBCs with low average complexity. This algorithm can be modified to implement both the ACZF and ACZF-SIC decoders. The ACZF-SIC decoder can be implemented with only a minor modification, and we explain this below.

We consider STBCs that are encoded by a square $M$-QAM constellation. In this case the real and imaginary parts of each complex symbol $s_i$ are encoded independently by a $\sqrt{M}$-ary PAM constellation. Thus we use a real sphere decoder to implement the ACZF-SIC decoder. When the constellation $\mathcal{A} \subset \mathbb{C}$ is arbitrary one can use the complex sphere decoder with the same modifications as explained below.

Let $m = \arg\max_{\ell \in \{1,\ldots,L\}} \det(\mathbf{G}_{\mathcal{I}_\ell}^H \mathbf{G}_{\mathcal{I}_\ell})$. Then we have

$$\mathbf{y} = \sqrt{\mathsf{SNR}}[\mathbf{G}_{\mathcal{I}_m} \ \mathbf{G}_{\mathcal{I}_m^c}] \begin{bmatrix} \mathbf{s}_{\mathcal{I}_m} \\ \mathbf{s}_{\mathcal{I}_m^c} \end{bmatrix} + \mathbf{n}. \quad (15)$$

Represent $\sqrt{\mathsf{SNR}}[\mathbf{G}_{\mathcal{I}_m} \ \mathbf{G}_{\mathcal{I}_m^c}]$ by $\mathbf{G}_0$ and $[\mathbf{s}_{\mathcal{I}_m}^T \ \mathbf{s}_{\mathcal{I}_m^c}^T]^T$ by $\mathbf{s}_0$, then decoding $\mathbf{s}$ is equivalent to decoding $\mathbf{s}_0$. For any complex vector $\mathbf{u}$, let $\check{\mathbf{u}}$ be the real vector obtained from $\mathbf{u}$ by replacing every element of $u_i$ of $\mathbf{u}$ by the tuple $[(u_i)_{Re} \ (u_i)_{Im}]^T$. Let $\check{\mathbf{G}}_0$ be the $2N_rT \times 2K$ real matrix obtained from the matrix $\mathbf{G}_0 = [g_{i,j}]$ by replacing every element $g_{i,j}$ by the $2 \times 2$ matrix

$$\begin{bmatrix} (g_{i,j})_{Re} & -(g_{i,j})_{Im} \\ (g_{i,j})_{Im} & (g_{i,j})_{Re} \end{bmatrix}.$$

Then (15) is equivalent to the real system $\check{\mathbf{y}} = \check{\mathbf{G}}_0 \check{\mathbf{s}}_0 + \check{\mathbf{n}}$. The ACZF-SIC algorithm decodes the first $2\lambda$ symbols of $\check{\mathbf{s}}_0$ with ZF-SIC by conditioning on the remaining $2(K-\lambda)$ symbols. Note that the elements of $\check{\mathbf{s}}_0$ are encoded with a $\sqrt{M}$-ary PAM signal set with centroid at zero. With a suitable scaling and translation this system can be modified into

$$\tilde{\mathbf{y}} = \check{\mathbf{G}}_0 \mathbf{x} + \mathbf{n},$$

where the entries of the $2K$ dimensional real symbol vector $\mathbf{x}$ are encoded with the alphabet $\mathbb{Z}_{\sqrt{M}} = \{0, 1, \ldots, \sqrt{M}-1\}$. Let the QR decomposition of $\check{\mathbf{G}}_0$ be $[\mathbf{Q} \ \mathbf{Q}']\begin{bmatrix} \mathbf{R} \\ \mathbf{0} \end{bmatrix}$, where $\mathbf{R} = [r_{i,j}]$ is $2K \times 2K$ upper triangular matrix and $\mathbf{Q}$ is a $2N_rT \times 2K$ matrix. Searching for a lattice point within a squared distance of $C$ from $\tilde{\mathbf{y}}$ is equivalent to searching for a vector $\mathbf{x} \in \mathbb{Z}_{\sqrt{M}}$ such that

$$||\mathbf{y}' - \mathbf{R}\mathbf{x}||^2 \leq C',$$

where $\mathbf{y}' = \mathbf{Q}^T \tilde{\mathbf{y}}$ and $C' = C - ||\mathbf{Q}'^T \tilde{\mathbf{y}}||^2$ [27].

The sphere-decoding implementation of ACZF-SIC decoder with Schnorr-Euchner enumeration [28] is given in Algorithm 2. This algorithm is a modification of Algorithm II of [27]. The variable $i$ represents the current stage of the sphere decoder, $T_i$ is the accumulated Euclidean distance of the current lattice point from $\mathbf{y}'$ at stage $i-1$, and $\xi_i$ represents the interference faced by $x_i$ from the already detected symbols $x_{i+1}, \ldots, x_{2K}$. The variable $d$ is the square of the current search radius, and is initialized with the value of $\infty$. Steps 2 and 6 together detect the symbols in the Schnorr-Euchner order for $i = 2\lambda+1, \ldots, 2K$. However, if $i \leq 2\lambda$ the symbol $x_i$ is ZF-SIC decoded, i.e., it is set to the nearest element in $\mathbb{Z}_{\sqrt{M}}$ after removing the interference from already detected symbols. If the symbol $x_i$ is within the search sphere and constellation boundaries, then Step 3 updates the values of $T_{i-1}$, $\xi_{i-1}$ and moves to the next stage, i.e., the $(i-1)^{th}$ stage. If, however,

**Input**: $\mathbf{y}'$ and $\mathbf{R}$
**Output**: Decoded symbol vector $\hat{\mathbf{x}}$.
%% Initialization
**Step 1**. Set $i := 2K$, $T_{2K} := 0$, $\xi_{2K} := 0$, $d := \infty$.

**Step 2**. **if** $i \leq 2\lambda$ **then**
$\quad x_i := \min\{\sqrt{M} - 1, \max\{0, \mathsf{rnd}\left(\frac{y'_i - \xi_i}{r_{i,i,}}\right)\}\}$
**end**
**else**
$\quad x_i := \mathsf{rnd}\left(\frac{y'_i - \xi_i}{r_{i,i,}}\right)$, $\Delta_i := \mathsf{sign}(y'_i - \xi_i - r_{i,i}x_i)$.
**end**
Go to Step 3.

**Step 3**. **if** $d < T_i + |y'_i - \xi_i - r_{i,i}x_i|^2$ **then**
$\quad$ %% We are outside the sphere.
$\quad$ Go to Step 4.
**end**
**else if** $x_i \leq -1$ or $x_i \geq \sqrt{M}$ **then**
$\quad$ %% We are outside the constellation boundaries.
$\quad$ Go to Step 6.
**end**
**else**
$\quad$ %% We are inside the sphere and
$\quad$ %% the constellation boundaries
$\quad$ **if** $i > 1$ **then**
$\quad\quad \xi_{i-1} := \sum_{j=i}^{2K} r_{i-1,j}x_j$,
$\quad\quad T_{i-1} = T_i + |y'_i - \xi_i - r_{i,i}x_i|^2$, $i := i-1$, go to Step 2.
$\quad$ **end**
$\quad$ **else**
$\quad\quad$ %% $i = 1$, a valid lattice point is found.
$\quad\quad$ Go to Step 5.
$\quad$ **end**
**end**

**Step 4**. **if** $i = 2K$ **then**
$\quad$ Terminate.
**end**
**else if** $i \leq 2\lambda$ **then**
$\quad i := 2\lambda + 1$, go to Step 6.
**end**
**else**
$\quad i := i + 1$, go to Step 6.
**end**

**Step 5**. Set $d := T_1 + |y'_1 - \xi_1 - r_{1,1}x_1|^2$, $\hat{\mathbf{x}} = \mathbf{x}$, $i := 2\lambda + 1$, go to Step 6.

**Step 6**. %% Schnorr-Euchner enumeration.
$x_i := x_i + \Delta_i$, $\Delta_i = -\Delta_i - \mathsf{sign}(\Delta_i)$, go to Step 3.

**Algorithm 2:** Sphere decoding implementation of the ACZF-SIC decoder.

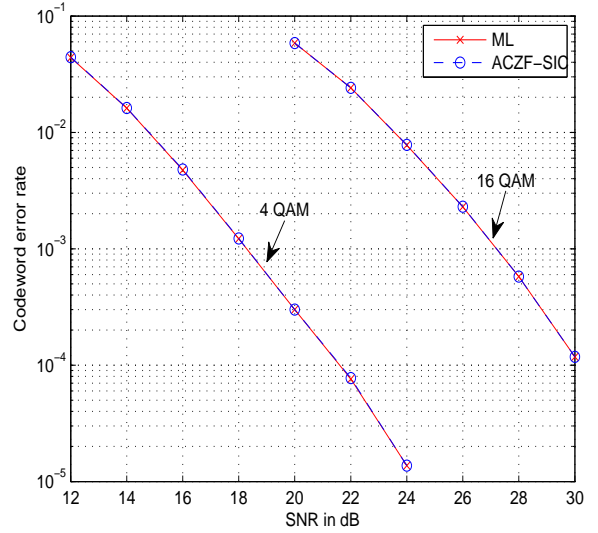

Fig. 1. Golden code, $N_r = 2$.

$i = 1$ then the decoder has found a new lattice point within the search sphere and in Step 5 the value of $\hat{\mathbf{x}}$ is updated, the new search radius is set to the distance of $\mathbf{R}\hat{\mathbf{x}}$ from $\mathbf{y}'$, and the decoder goes to the $i = 2\lambda+1$ stage. This is different from the usual sphere decoder where $i$ is set to 2 after a valid point has been found. This difference arises because in ACZF-SIC the symbols $x_1, \ldots, x_{2\lambda}$ are decoded by ZF-SIC, and hence require no Schnorr-Euchner enumeration. The case that the accumulated Euclidean distance from $\mathbf{y}'$ at the current stage exceeds $\sqrt{d}$ is handled in Step 4. In this case if the current stage is $2K$ then the program is terminated as no lattice point at a distance $\sqrt{d}$ or less can be found, else if $i > 2\lambda$ then the decoder goes to stage $i+1$ for Schnorr-Euchner enumeration, and if $i < 2\lambda$ it goes to stage $2\lambda + 1$ as the first $2\lambda$ symbols do not undergo enumeration.

## VI. SIMULATION RESULTS

The simulation results comparing the performance of the ACZF-SIC decoder with the ML decoder for the Golden code, 3 antenna TAST code, 4 antenna Perfect code and the Srinath-Rajan code are given in Fig. 1- 4, in that order. In all the cases it can be seen that ACZF-SIC performs identical to the ML decoder. The simulations provided in [15] show that the same holds true for the 3 antenna Perfect code with ACZF decoding.

From Table I we see that for each of these five codes the ACZF-SIC complexity is strictly less than the ML decoding complexity. From the table we also see that the ACZF-SIC complexity of these codes is less than or equal to the complexity of the full-diversity codes with least known ML decoding complexity. Thus the decoding complexity of the best performing codes for $2, 3$ and $4$ antennas with low ML decoding complexity (complexity less than $M^K$) can be reduced further by using the ACZF-SIC decoder without trading off the error performance.

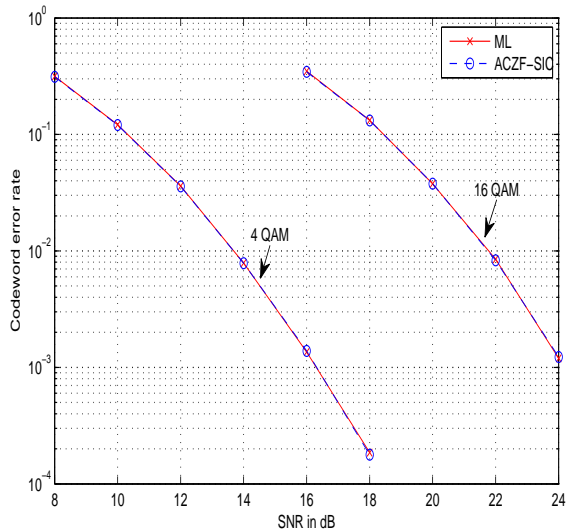

Fig. 2. The $3 \times 3$ TAST code, $N_r = 3$.

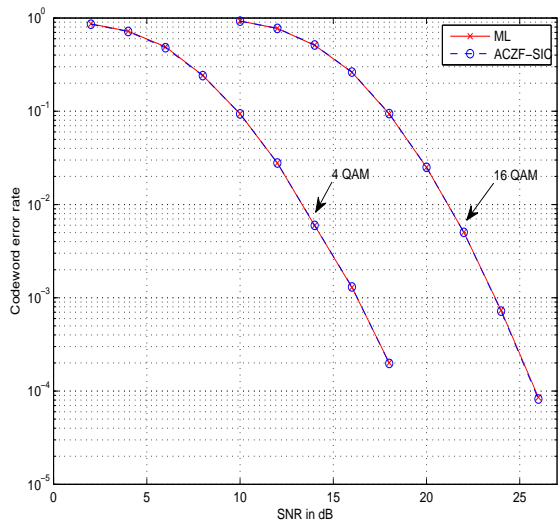

Fig. 4. The Srinath-Rajan code, $N_r = 2$.

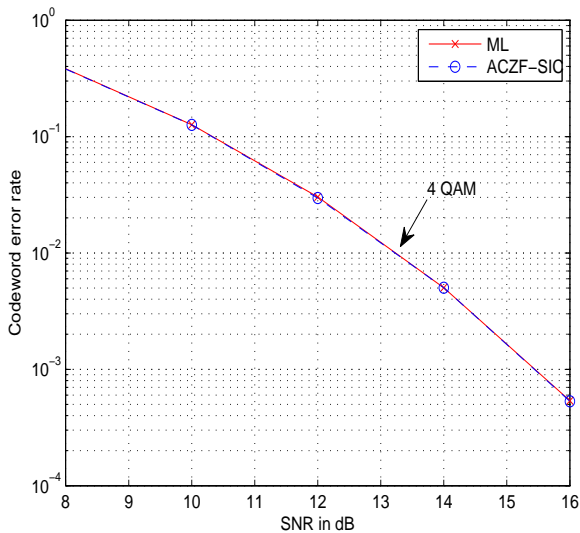

Fig. 3. The 4 antenna Perfect code, $N_r = 4$.

## VII. CONCLUSION

In this paper, we have introduced two new low complexity decoding techniques for STBCs - the ACZF and ACZF-SIC decoders. We have given sufficient conditions for an STBC to give full-diversity with ACZF/ACZF-SIC decoding, and shown that the Golden code, 3 and 4 antenna Perfect code, 3 antenna TAST code and the Srinath-Rajan code can be full-diversity ACZF/ACZF-SIC decoded with complexity less than that of ML decoding. Simulations show that this advantage in decoding comfort comes with no loss in error performance with respect to ML decoding. These five codes along with ACZF/ACZF-SIC decoding outperform all known codes for $N_t \leq 4$ in terms of complexity and error performance. The following problems are yet to be settled.

- The reason for the essentially-ML performance of ACZF/ACZF-SIC decoders is to be investigated. Is there a criterion that ensures that a given non-ML decoding technique has error performance close to that of the ML decoder?
- Constructing/Identifying STBCs for $N_t > 4$ that allow full-diversity ACZF/ACZF-SIC decoding. For example, are all TAST codes [19] or all codes from Cyclic Division Algebras [29] full-diversity ACZF/ACZF-SIC decodable?

## APPENDIX A
## PROOF OF THEOREM 1

In order to prove the theorem we first derive a lower bound for $\det(\mathbf{G}_{\mathcal{I}_m}^H \mathbf{G}_{\mathcal{I}_m})$, we then use this result to derive an upper bound on the probability of decoding error.

Assume that the hypothesis of the theorem is true. Let $\mathbf{h} = \text{vec}(\mathbf{H})$. For every $\ell = 1, \ldots, L$ each entry of $\mathbf{G}_{\mathcal{I}_\ell}$ is a complex linear combination of entries of $\mathbf{h}$. Thus the function $f_\ell(\mathbf{h}) = \det(\mathbf{G}_{\mathcal{I}_\ell}^H \mathbf{G}_{\mathcal{I}_\ell})$ is a polynomial in entries of $\mathbf{h}$ and hence is a continuous function. Note that $m$ itself is a function of $\mathbf{h}$, and

$$\det(\mathbf{G}_{\mathcal{I}_m}^H \mathbf{G}_{\mathcal{I}_m}) = f(\mathbf{h}) = \max_{\ell \in \{1,\ldots,L\}} f_\ell(\mathbf{h})$$

is a also a continuous function since it is the maximum of $L$ individual continuous functions. We will now follow an argument similar to [1], [2] to derive a lower bound on $f$. Consider the unit sphere $\mathcal{S}$ in $\mathbb{C}^{N_r T}$. From the hypothesis of the theorem, $f(\mathbf{h}) > 0$ for every $\mathbf{h} \in \mathcal{S}$. Since $\mathcal{S}$ is a compact set and $f$ is continuous, there exists a real number $c > 0$ such that $f(\mathbf{h}) \geq c$ for any $\mathbf{h} \in \mathcal{S}$. Since each $\mathbf{G}_{\mathcal{I}_\ell}$, $\ell = 1, \ldots, L$ is a linear function of $\mathbf{h}$ it satisfies $\mathbf{G}_{\mathcal{I}_\ell}(\mathbf{h}) = ||\mathbf{h}|| \ \mathbf{G}_{\mathcal{I}_\ell}\left(\frac{\mathbf{h}}{||\mathbf{h}||}\right)$. Therefore $f_\ell(\mathbf{h}) = ||\mathbf{h}||^{2\lambda} f_\ell\left(\frac{\mathbf{h}}{||\mathbf{h}||}\right)$ for all $\ell = 1, \ldots, L$, and hence $f(\mathbf{h}) = ||\mathbf{h}||^{2\lambda} f\left(\frac{\mathbf{h}}{||\mathbf{h}||}\right)$ for any $\mathbf{h} \neq bf0$. Since $\frac{\mathbf{h}}{||\mathbf{h}||} \in \mathcal{S}$ we have $f\left(\frac{\mathbf{h}}{||\mathbf{h}||}\right) \geq c$. Thus,

$$\det(\mathbf{G}_{\mathcal{I}_m}^H \mathbf{G}_{\mathcal{I}_m}) = f(\mathbf{h}) \geq c \, ||\mathbf{h}||^{2\lambda}. \quad (16)$$

Assume that the transmitted information vector is $\mathbf{b} \in \mathcal{A}^K$. Let $\mathcal{E}_1$ be the event that the tuple $(\mathbf{b}_{\mathcal{I}_m^c}, \mathbf{b}_{\mathcal{I}_m})$ does not belong to the set of vectors (3) obtained at the end of the second step of the decoding process. Let $\mathcal{E}$ denote the event of the decoder deciding in favour of the wrong codeword. Then

$$\begin{aligned} \mathsf{P}(\mathcal{E}) &= \mathsf{P}(\mathcal{E} \cap \mathcal{E}_1) + \mathsf{P}(\mathcal{E} \cap \mathcal{E}_1^c) \\ &\leq \mathsf{P}(\mathcal{E}_1) + \mathsf{P}(\mathcal{E}/\mathcal{E}_1^c). \end{aligned} \quad (17)$$

We will now upper bound each of the two terms in the above expression to complete the proof.

Since $\mathbf{b}$ is the transmitted vector we have $\mathbf{y} = \sqrt{\mathsf{SNR}}(\mathbf{G}_{\mathcal{I}_m^c}\mathbf{b}_{\mathcal{I}_m^c} + \mathbf{G}_{\mathcal{I}_m}\mathbf{b}_{\mathcal{I}_m}) + \mathbf{n}$. The vector $\hat{\mathbf{s}}_{\mathcal{I}_m}(\mathbf{b}_{\mathcal{I}_m^c})$ which belongs to the set of vectors in (3) is the output of ZF decoder for the system

$$\begin{aligned} \mathbf{y}(\mathbf{b}_{\mathcal{I}_m^c}) &= \mathbf{y} - \sqrt{\mathsf{SNR}} \mathbf{G}_{\mathcal{I}_m^c} \mathbf{b}_{\mathcal{I}_m^c} \\ &= \sqrt{\mathsf{SNR}} \mathbf{G}_{\mathcal{I}_m} \mathbf{b}_{\mathcal{I}_m} + \mathbf{n}. \end{aligned}$$

This is a space-time block coded system with a ZF receiver whose equivalent channel matrix is $\mathbf{G}_{\mathcal{I}_m}$. The full-diversity (i.e., order of $N_t N_r$ diversity) criterion for this system is given in Theorem 1 of [1], and it coincides with (16). Following the same steps as in the proof of Theorem 1 of [1] we can show that

$$\mathsf{P}(\mathcal{E}_1) \leq \mathsf{P}\left(\hat{\mathbf{s}}_{\mathcal{I}_m}(\mathbf{b}_{\mathcal{I}_m^c}) \neq \mathbf{b}_{\mathcal{I}_m}\right) \leq c_1 \mathsf{SNR}^{-N_t N_r}, \quad (18)$$

for some real number $c_1 > 0$.

The term $\mathsf{P}(\mathcal{E}/\mathcal{E}_1^c)$ is the probability that the minimum distance decoder does not decide in favour of $\mathbf{b}$ given that the information symbol vector $(\mathbf{b}_{\mathcal{I}_m^c}, \mathbf{b}_{\mathcal{I}_m})$ does belong to the set (3). This step of the decoder is same as that of ML decoding of $\mathcal{C}$ except that the search space of the ML decoder has been reduced from the entire codebook $\mathcal{C}$ of size $M^K$ to the set of $M^{K-\lambda}$ codewords corresponding to the information vectors in (3). Since the transmitted codeword belongs to the reduced set of codewords, the probability that the wrong codeword is output at this step is upper bounded by the probability of error of the ML decoder of $\mathcal{C}$. Thus there exists a constant $c_2 > 0$ such that

$$\mathsf{P}(\mathcal{E}/\mathcal{E}_1^c) \leq c_2 \mathsf{SNR}^{-rN_r}, \quad (19)$$

where $r = \min\{rank(\mathbf{X} - \mathbf{X}') \mid \mathbf{X}, \mathbf{X}' \in \mathcal{C}, \mathbf{X} \neq \mathbf{X}'\}$. Since $r \leq N_t$, from (17), (18) and (19) we have $\mathsf{P}(\mathcal{E}) \leq (c_1 + c_2)\mathsf{SNR}^{-rN_r}$ at high signal to noise ratios. Thus the ACZF decoder achieves the same diversity order as the ML decoder.

## APPENDIX B
## PROOF OF THEOREM 2

We will first prove that for a given $\ell = 1, \ldots, L$, the matrix $\mathbf{G}_{\mathcal{I}_\ell}$ has full column rank if and only if $\mathbf{X}_{\mathcal{I}_\ell}(\mathbf{u})\mathbf{H} \neq \mathbf{0}$ for any non-zero vector $\mathbf{u} = [u_1 \ u_2 \ \cdots \ u_\lambda] \in \mathbb{C}^\lambda$. Let $\mathcal{I}_\ell = \{i_1, i_2, \ldots, i_\lambda\}$ i.e.,

$$\mathbf{G}_{\mathcal{I}_\ell} = [\text{vec}(\mathbf{A}_{i_1}\mathbf{H}) \ \text{vec}(\mathbf{A}_{i_2}\mathbf{H}) \cdots \text{vec}(\mathbf{A}_{i_\lambda}\mathbf{H})],$$

The columns of $\mathbf{G}_{\mathcal{I}_\ell}$ are linearly independent if and only if the matrices $\mathbf{A}_{i_1}\mathbf{H}, \ldots, \mathbf{A}_{i_\lambda}\mathbf{H}$ are linearly independent, i.e., the matrix $\sum_{j=1}^{\lambda} u_j \mathbf{A}_{i_j} \mathbf{H} \neq \mathbf{0}$ for any non-zero vector $\mathbf{u}$. This is same as $\mathbf{X}_{\mathcal{I}_\ell}(\mathbf{u})\mathbf{H} \neq \mathbf{0}$ for any non-zero vector $\mathbf{u}$.

Using the result from the previous paragraph we will now show that the criteria of Theorems 1 and 2 are equivalent. Let $\mathcal{E}_A$ denote the event that the criterion of Theorem 1 is satisfied, and let $\mathcal{E}_B$ be the event corresponding to Theorem 2. We need to prove that $\mathcal{E}_A$ implies $\mathcal{E}_B$ and vice-versa, or equivalently $\mathcal{E}_A^c$ implies $\mathcal{E}_B^c$ and vice-versa. The event $\mathcal{E}_A$ is that for every $\mathbf{H} \neq \mathbf{0}$, at least one the $L$ matrices $\{\mathbf{G}_{\mathcal{I}_\ell}\}$ is of full column rank. From the result in the previous paragraph this happens if and only if for any given $\mathbf{H} \neq \mathbf{0}$ there exists an $\ell \in \{1, \ldots, L\}$ such that $\mathbf{X}_{\mathcal{I}_\ell}(\mathbf{u}_\ell)\mathbf{H} \neq \mathbf{0}$ for any $\mathbf{u}_\ell \neq \mathbf{0}$.

Suppose $\mathcal{E}_A^c$ is true. There exists an $\mathbf{H} \neq \mathbf{0}$ and non-zero vectors $\mathbf{u}_1, \ldots, \mathbf{u}_L$ such that $\mathbf{X}_{\mathcal{I}_\ell}(\mathbf{u}_\ell)\mathbf{H} = \mathbf{0}$ for $\ell = 1, \ldots, L$. Thus all $\mathbf{X}_{\mathcal{I}_\ell}(\mathbf{u}_\ell)$ have a common non-zero nullspace, and hence the matrix $\widetilde{\mathbf{X}}$ in (6) does not have full column rank. Thus $\mathcal{E}_B^c$ is true.

Now suppose $\mathcal{E}_B^c$ is true. This means there exist non-zero vectors $\mathbf{u}_1, \ldots, \mathbf{u}_L$ such that $\widetilde{\mathbf{X}}$ has a non-zero nullspace. Choose $\mathbf{H} \neq \mathbf{0}$ to be any matrix with columns in the nullspace of $\widetilde{\mathbf{X}}$. Thus, $\widetilde{\mathbf{X}}\mathbf{H} = \mathbf{0}$ and hence $\mathbf{X}_{\mathcal{I}_\ell}(\mathbf{u}_\ell)\mathbf{H} = \mathbf{0}$ for each $\ell = 1, \ldots, L$ for this choice of non-zero vectors $\mathbf{u}_1, \ldots, \mathbf{u}_L$. This negates the event $\mathcal{E}_A$. This completes the proof.

## APPENDIX C
## PROOF OF LEMMA 1

The criteria of Theorems 1 and 2 are equivalent, thus it is enough to prove that the criterion of Theorem 1 implies full-diversity with ACZF-SIC decoding. The proof is similar to that of Theorem 1. The difference lies in upper bounding the term $\mathsf{P}(\mathcal{E}_1)$, which itself is upper bounded by the probability of error of the ZF-SIC decoder for the system

$$\mathbf{y}(\mathbf{b}_{\mathcal{I}_m^c}) = \sqrt{\mathsf{SNR}} \mathbf{G}_{\mathcal{I}_m} \mathbf{b}_{\mathcal{I}_m} + \mathbf{n}.$$

Let $\mathcal{I}_m = \{i_1, \ldots, i_\lambda\}$ and let $\mathsf{SNR}_k$ and $\mathsf{SNR}'_k$ be the post-processing signal to noise ratios for the symbol $s_{i_k}$ for ZF and ZF-SIC decoders respectively. Further, let $\mathsf{P}_k$ and $\mathsf{P}'_k$ be the probability of error for $s_{i_k}$ with ZF and ZF-SIC decoder respectively.

Suppose the criterion of Theorem 1 is satisfied. From the proof of Theorem 1 we have that $\mathsf{P}_k \leq c_k \mathsf{SNR}^{-N_t N_r}$ at high SNR, for some real number $c_k > 0$ for $k = 1, \ldots, \lambda$. We will now show that $\mathsf{P}'_k$ is of the order of $\mathsf{SNR}^{-N_t N_r}$ for $k = 1, \ldots, \lambda$. Given this result on $\mathsf{P}'_k$ the rest of the proof is similar to that of Theorem 1.

Now let the ZF-SIC decoder decode the symbols in the order $i_\lambda, i_{\lambda-1}, \ldots, i_1$, then we have $\mathsf{SNR}_\lambda = \mathsf{SNR}'_\lambda$ for every channel $\mathbf{H}$. Hence $\mathsf{P}'_\lambda = \mathsf{P}_\lambda = O(\mathsf{SNR}^{-N_t N_r})$. Now consider the ZF-SIC detection of the symbol $s_{i_{\lambda-1}}$. If the symbol $s_{i_\lambda}$ is decoded correctly the number of interferences faced by $s_{i_{\lambda-1}}$ is one less in ZF-SIC receiver than in the ZF receiver. In that case the post-processing signal to noise ratio $\mathsf{SNR}'_{\lambda-1} \geq \mathsf{SNR}_{\lambda-1}$ for any channel $\mathbf{H}$. Therefore if $s_{i_\lambda}$ were decoded correctly by the ZF-SIC receiver then $\mathsf{P}'_{\lambda-1} \leq \mathsf{P}_{\lambda-1} = O(\mathsf{SNR}^{-N_t N_r})$. Now the overall probability of error for $s_{i_{\lambda-1}}$ is upper bounded by the sum of probability of error for $s_{i_\lambda}$ and the probability of detecting $s_{i_{\lambda-1}}$ wrongly given that $s_{i_\lambda}$ was detected correctly. Since both these terms are of the order of $\mathsf{SNR}^{-N_t N_r}$ we have that $\mathsf{P}'_{\lambda-1} = O(\mathsf{SNR}^{-N_t N_r})$. Using a similar argument we can prove that $\mathsf{P}'_k = O(\mathsf{SNR}^{-N_t N_r})$ for all $k = 1, \ldots, \lambda$.